\documentclass[letterpaper, english, aps, prl, reprint, twocolumn, superscriptaddress]{revtex4-2}
\usepackage{amsmath}
\usepackage{amssymb}
\usepackage{graphicx}
\usepackage{dcolumn}
\usepackage{bm}
\usepackage[colorlinks=true,linkcolor=blue,anchorcolor=blue, citecolor=cyan,urlcolor=cyan]{hyperref}
\usepackage[mathlines]{lineno}
\usepackage{ulem}
\usepackage{epstopdf}
\newcommand{\ket}[1]{|#1 \rangle}
\newcommand{\bra}[1]{\langle #1 |}
\begin{document}
\title{Two-dimensional fully-compensated Ferrimagnetism}
\author{Yichen Liu}
\thanks{These authors contributed equally to this work.}
\affiliation{Centre for Quantum Physics, Key Laboratory of Advanced Optoelectronic Quantum Architecture and Measurement (MOE), School of Physics, Beijing Institute of Technology, Beijing 100081, China}
\author{San-Dong Guo}
\thanks{These authors contributed equally to this work.}
\affiliation{School of Electronic Engineering, Xi'an University of Posts and Telecommunications, Xi'an 710121, China}
\author{Yongpan Li}
\affiliation{Centre for Quantum Physics, Key Laboratory of Advanced Optoelectronic Quantum Architecture and Measurement (MOE), School of Physics, Beijing Institute of Technology, Beijing 100081, China}
\author{Cheng-Cheng Liu}
\email{ccliu@bit.edu.cn}
\affiliation{Centre for Quantum Physics, Key Laboratory of Advanced Optoelectronic Quantum Architecture and Measurement (MOE), School of Physics, Beijing Institute of Technology, Beijing 100081, China}
\begin{abstract}
Antiferromagnetic spintronics has long been a subject of intense research interest, and the recent introduction of altermagnetism has further ignited enthusiasm in the field. However, fully-compensated ferrimagnetism, which exhibits band spin splitting but zero net magnetization, has yet to receive enough attention. Since the experimental preparation of two-dimensional (2D) magnetic van der Waals (vdW) materials in 2017, 2D magnetic materials, thanks to their super tunability, have quickly become an important playground for spintronics. Here, we extend the concept of fully-compensated ferrimagnetism (fFIM) to two dimensions and propose 2D \textit{filling-enforced} fFIM, demonstrate its stability and ease of manipulation, and present three feasible realization schemes with respective exemplary candidate materials. A simple model for 2D fully-compensated ferrimagnets (fFIMs) is developed. Further investigation of 2D fFIMs' physical properties reveals that they not only exhibit significant magneto-optical response but also show fully spin-polarized currents and the anomalous Hall effect in the half-metallic states, displaying characteristics previously almost exclusive to ferromagnetic materials, greatly broadening the research and application prospects of spintronic materials.
\end{abstract}
\maketitle
\textit{Introduction.---} Traditionally, collinear magnetism can be classified into ferromagnetism, antiferromagnetism, and ferrimagnetism. In ferromagnets, all magnetic moments are aligned in the same direction with non-zero magnetization, and the bands are spin-splitting. In conventional antiferromagnets (cAFMs), the magnetic moments are arranged antiparallel with zero net magnetization, and the bands are spin-degenerate. However, the spin-degenerate nature of the energy bands in cAFMs eliminates the anomalous Hall effect (AHE) and magneto-optical effect, thereby restricting their potential applications~\cite{baltzAntiferromagneticSpintronics2018}. Ferrimagnetism has antiparallel magnetic moments that usually cannot completely cancel each other out. Recently, a new type of magnetic material, the so-called altermagnets, with zero net magnetization and anisotropic spin-splitting bands, was proposed\cite{smejkalEmergingResearchLandscape2022,mazinEditorialAltermagnetismNew2022}. Compared to ferromagnetic materials, magnetic materials with null magnetization, such as cAFMs and altermagnets, offer several advantages, including ultrafast dynamics, the absence of stray fields, and enhanced stability in the presence of magnetic fields~\cite{baltzAntiferromagneticSpintronics2018}. However, this classification overlooks a significant class of zero net magnetization, the fully-compensated ferrimagnetism~\cite{van1995half,Akai2006prl,wurmehlValenceElectronRules2006,mazinEditorialAltermagnetismNew2022}, which is firstly issued by de Groot et al.~\cite{van1995half}. Unlike cFAMs and altermagnets, fully-compensated ferrimagnets (fFIMs) achieve zero net magnetization through appropriate filling instead of symmetry.
\begin{figure}[t]
    \centering
    \includegraphics[width=0.48\textwidth]{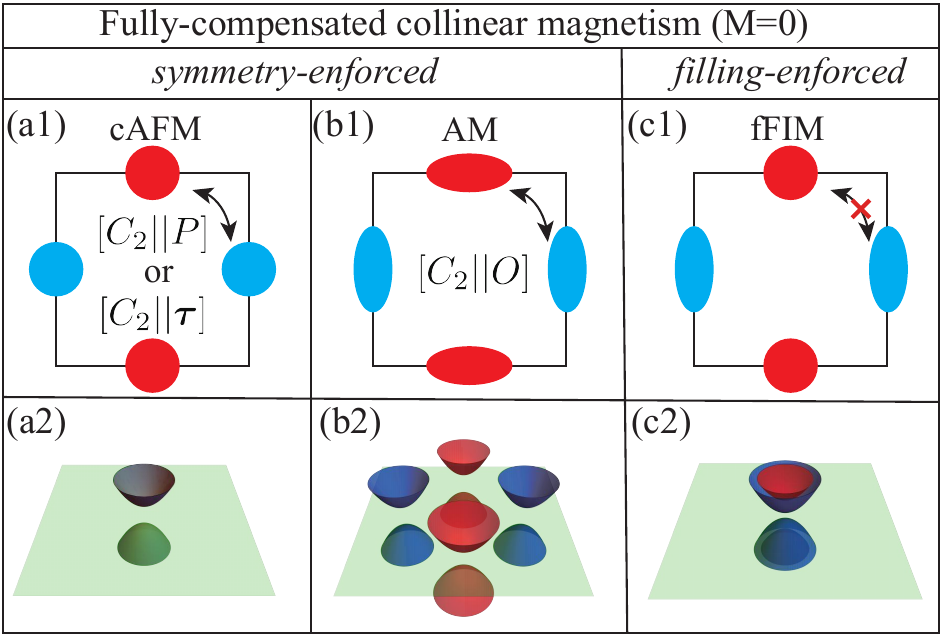}
    \caption{\label{fig:illu}Three categories of fully-compensated collinear magnetism with null net magnetization (M=0). (a1) Conventional antiferromagnetism (cAFM) is characterized by magnetic sublattices that are interconnected by $[C_2||P]$ and $[C_2||\tau]$, featuring Kramers degeneracy, as shown in (a2). Here $C_2$, $P$, and $\tau$ are the two-fold rotation perpendicular to the spin axis, inversion, and translation, respectively. (b1) Two magnetic sublattices in altermagnetism (AM) are connected by $[C_2||O]$, where $O$ represents a rotation or mirror symmetry. This results in anisotropic spin splitting, as depicted in (b2). (c1) Fully-compensated ferrimagnetism (fFIM), in which the two magnetic sublattices have no symmetry connection, exhibits isotropic spin splitting (c2). The zero net magnetization in cAFM and AM is \textit{symmetry-enforced}, while in fFIM is \textit{filling-enforced}.  In (a2)-(c2), the gray surfaces represent spin-degenerate bands, the red and blue surfaces correspond to spin-up and spin-down bands, respectively, and the green plane indicates the Fermi level.}
\end{figure}

Previous studies on fFIMs have been limited to three-dimensional (3D) systems with sophisticated crystal structures, such as the double perovskite oxides, Heusler system, and organic compound~\cite{Pickett1998_prb,Hu2008prl,siewierskaMagneticOrderMagnetotransport2021,jamerCompensatedFerrimagnetismZeroMoment2017,zicDesigningFullyCompensated2016,Coey2002jap,hu2012half,ozdoganAbinitioInvestigationElectronic2015,stinshoffCompletelyCompensatedFerrimagnetism2017,fleischerMagnetoopticKerrEffect2018,kawamuraCompensatedFerrimagnetsColossal2024}. Moreover, clear experimental evidence for fFIMs remains elusive. Since the experimental fabrication of CrI$_3$\cite{huangLayerdependentFerromagnetismVan2017} and Cr$_2$Ge$_2$Te$_6$\cite{gongDiscoveryIntrinsicFerromagnetism2017c} in 2017,  2D magnetic van der Waals (mvdW) materials with excellent adjustability have immediately attracted tremendous interest. So far, a richer variety of 2D mvdW material systems have been synthesized\cite{dengGatetunableRoomtemperatureFerromagnetism2018,bonillaStrongRoomtemperatureFerromagnetism2018a,kleinControlStructureSpin2022}, including Fe$_3$Ge$_2$Te$_4$, CrSBr, and VSe$_2$, etc.
Considering the abundance and excellent tunability of 2D magnetic materials and the fact that fFIMs have the advantage of combining the band spin splitting of ferromagnetism with a zero net magnetization in cAFMs but lack of clear experimental confirmation, extending fully-compensated ferrimagnetism to two dimensions is highly necessary and very attractive.

In the Letter, we conceptualize the 2D \textit{filling-enforced} fully-compensated ferrimagnetism (fFIM). A simple tight-binding model is developed, and by symmetry arguments, the transition between 2D \textit{filling-enforced} fFIMs, cAFMs, and altermagnets is shown. Based on theoretical analysis, we propose three feasible approaches to constructing fFIMs using 2D mvdW materials: Constructing Janus structure, adding staggered potential, and using element substitution or alloying, each with exemplary candidate materials. Despite having null net magnetization, we find that fFIMs can exhibit remarkable magneto-optical effects, such as observable Kerr and Faraday angles, AHE, and fully spin-polarized currents. Taking classic 2D mvdW materials CrI$_3$ and half-metallic YI$_2$~\cite{huangLargeSpontaneousValley2023} as examples, we show that 2D fFIMs not only readily implement experimentally but also display significant Kerr and Faraday angles and anomalous Hall conductivity (AHC).

\textit{Filling-enforced fully-compensated ferrimagnetism.---}
In the two kinds of fully-compensated collinear magnetism, i.e., conventional antiferromagnetism (cAFM) and altermagnetism, the magnetic sublattices are connected by symmetry, such as through the spatial inversion symmetry ($P$) in cAFM and through rotational or mirror symmetry ($O$) in altermagnetism. We thus call cAFM and altermagnetism \textit{symmetry-enforced} fully-compensated collinear magnetism, as illustrated in Figs.~\ref{fig:illu}(a) and (b). In contrast, the magnetic sublattices in 2D \textit{filling-enforced} fFIMs cannot be connected through any symmetry, resulting in ferromagnetic-like spin splitting, as shown in Fig.~\ref{fig:illu}(c). The net magnetization in usual ferrimagnetism is given by
\begin{equation}
M=\mu_B(N_{\uparrow}-N_{\downarrow}),
\end{equation}
with $\mu_B$ the Bohr magneton, where $N_{\uparrow}$ and $N_{\downarrow}$ represent the number of occupied states in the spin-up and spin-down channels, respectively.
As long as any one of the spin channels is gapped, regardless of whether the other spin channel has a gap or not, which corresponds to insulating ferrimagnetic states and half-metallic ferrimagnetic states, the net magnetization must be an integer Bohr magneton. This is because the total number of electrons $N_e$ is an integer, and the number of electrons in the gapped channel is also an integer, so the number of electrons in the other spin channel is an integer, too. In this case, the two integers $N_\uparrow$ and $N_\downarrow$ are two good quantum numbers that can be used to label different ferrimagnets. When the system has an appropriate filling, i.e., $N_\uparrow=N_\downarrow$, the net magnetization is strictly zero, corresponding to the fully-compensated ferrimagnetism (fFIM), which we call \textit{filling-enforced} fFIM. As long as the gap is not closed, the three integers $N_e$, $N_\uparrow$, and $N_\downarrow$ will not change, and integers do not change continuously, so $N_\uparrow=N_\downarrow$, that is, the fully-compensated characteristic is strictly maintained under reasonable perturbations. The reasonable perturbations here mean that they will not close the gap. Such \textit{filling-enforced} fFIM is protected by the gap-guaranteed spin quantization in one spin channel, which is similar to how stable topological quantum states are protected by a bulk energy gap~\cite{haldane1988model,kane2005z}. Since the \textit{filling-enforced} fFIM is not limited by symmetry constraints, it remains highly robust under external fields, such as electric fields and stress, regardless of whether these fields break the symmetry.

\begin{figure}[t]
    \centering
    \includegraphics[width=0.48\textwidth]{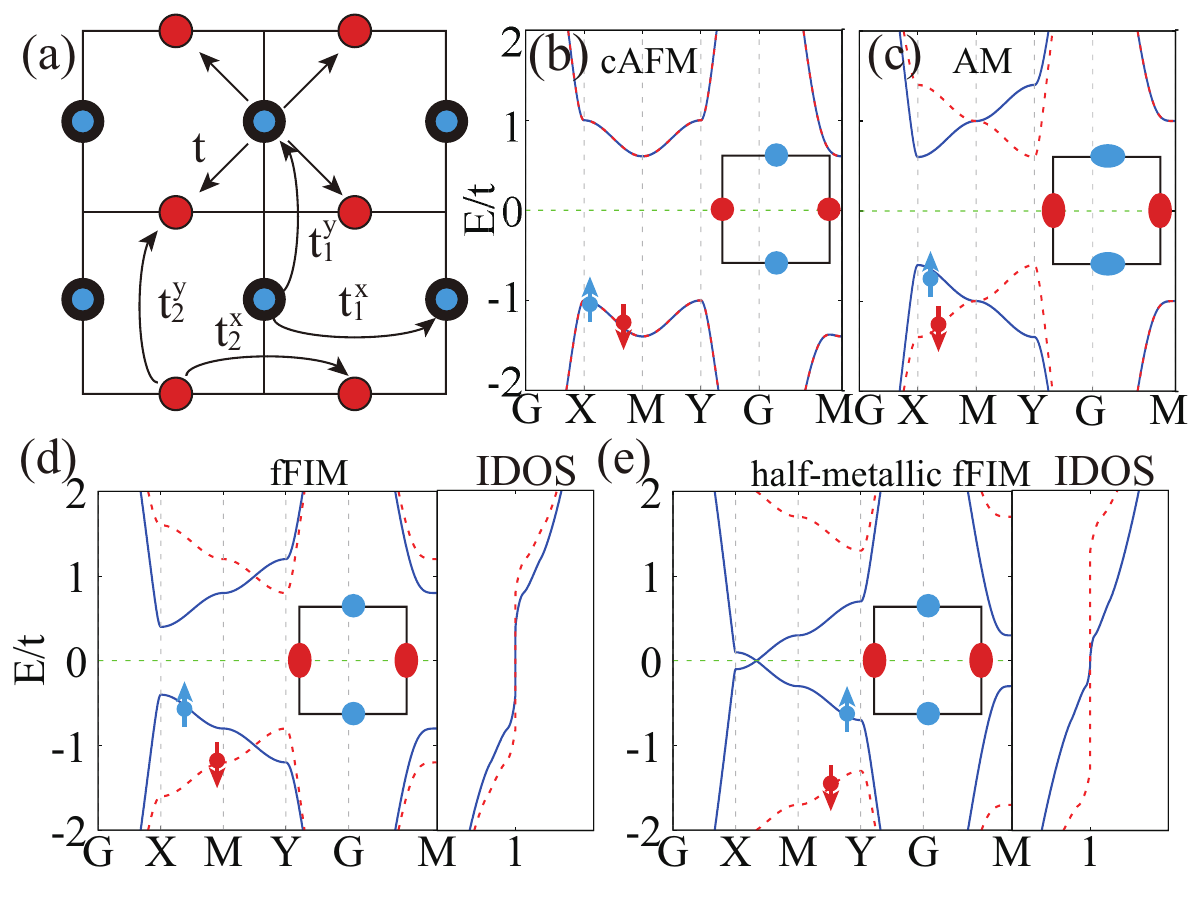}
    \caption{\label{fig:model} A simple model for the three categories of fully-compensated collinear magnetism, i.e., conventional antiferromagnetism (cAFM), altermagnetism (AM), and \textit{filling-enforced} fully-compensated ferrimagnetism (fFIM). (a) The illustration of our model. (b) When $\Delta=0$ and $t_{1}^{x/y}=t_{2}^{x/y}=0.1 t$, the system is cAFM with $[C_2||P]$ symmetry. (c) When $\Delta = 0$ and $t_{1/2}^{x/y}=-t_{1/2}^{y/x}=0.1 t$, the opposite spin sublattices are connected by $C_4$, transforming the system into AM with anistropic spin splitting. (d) (e) Setting $\Delta = 0.2t$ and $\Delta=0.7 t$ with $t_{1/2}^{x/y}=-t_{1/2}^{y/x}=0.1 t$ lift any symmetry that connects two sublattices, the system can be fFIM. Since the gap in one spin channel has never closed, the IDOSs of opposite spins are integer and remain equal at Fermi energy with $N_{\uparrow}=N_{\downarrow}$, resulting in a zero net magnetization. The half-metallic fFIM is shown in (e). $M= t$ is taken for (b)-(e). }
\end{figure}

We develop a simple 2D model to capture cAFMs, altermagnets, and \textit{filling-enforced} fFIMs and the transformation between them. The model is as follows
\begin{eqnarray}
H=&&t\sum_{i,d_j} (c_{1,i}^\dag c_{2,i+d_j}+ \text{h.c.})\nonumber\\
&&+\sum_{\alpha,i}(t_\alpha^x c_{\alpha,i}^\dag c_{\alpha,i+a_x}+t_\alpha^y c_{\alpha,i}^\dag c_{\alpha,i+a_y}+ \text{h.c.})\nonumber \\
&&+\Delta\sum_{\alpha,i} (-1)^\alpha c_{\alpha,i}^\dag c_{\alpha,i}+M\sum_{\alpha,i} (-1)^\alpha c_{\alpha,i}^\dag c_{\alpha,i} s_z,
\end{eqnarray}\label{Eq:model}
where $t$ represents the nearest-neighbor hopping, $\alpha=1,2$ denotes the sublattices located at $(0.5,0)$ and $(0,0.5)$, respectively, $t_{1}^{x/y}$ and $t_2^{x/y}$ are the next-nearest-neighbor hopping, $\Delta$ is the staggered potential, $s_z$ is the spin operator, and $M$ is the magnetization. $d_i$ and $a_{x/y}$ are the nearest vector and $x/y$ direction lattice vectors. The lattice structure is illustrated in Fig.~\ref{fig:model}(a). We use $t_{1}^{x/y}$, $t_2^{x/y}$, and $\Delta$ to regulate the symmetry of the system, corresponding to the following three cases of respective cAFMs, altermagnets, and \textit{filling-enforced} fFIMs.

When $\Delta=0$ and $t_{1}^{x/y}=t_{2}^{x/y}$, the sublattices with opposite spin can be connected through inversion $P$. The inversion center is located at (0.25,0.25), swapping $t_1^{x/y}$ with $t_2^{x/y}$. Since the two sublattices have opposite spin, the system possesses $[C_2||P]$ symmetry with $C_2$  a two-fold rotation perpendicular to the spin axis, resulting in Kramers degeneracy as illustrated in Fig.~\ref{fig:model}(b).

When $\Delta=0$ and $t_{1}^{x/y}\neq t_{2}^{x/y}$, the $[C_2||P]$ symmetry is broken. However, the sublattices with opposite magnetization can be connected through a $C_4$ rotation with the center at $(0, 0)$, resulting in $t_1^x \leftrightarrow t_2^y$ and $t_1^y \leftrightarrow t_2^x$. Thus, the system possesses $[C_2 || C_4]$ symmetry, inducing the hallmark anisotropic $d-$wave spin splitting in altermagnets, as depicted in Fig.~\ref{fig:model}(c).

\begin{figure}[t]
    \centering
    \includegraphics[width=0.47\textwidth]{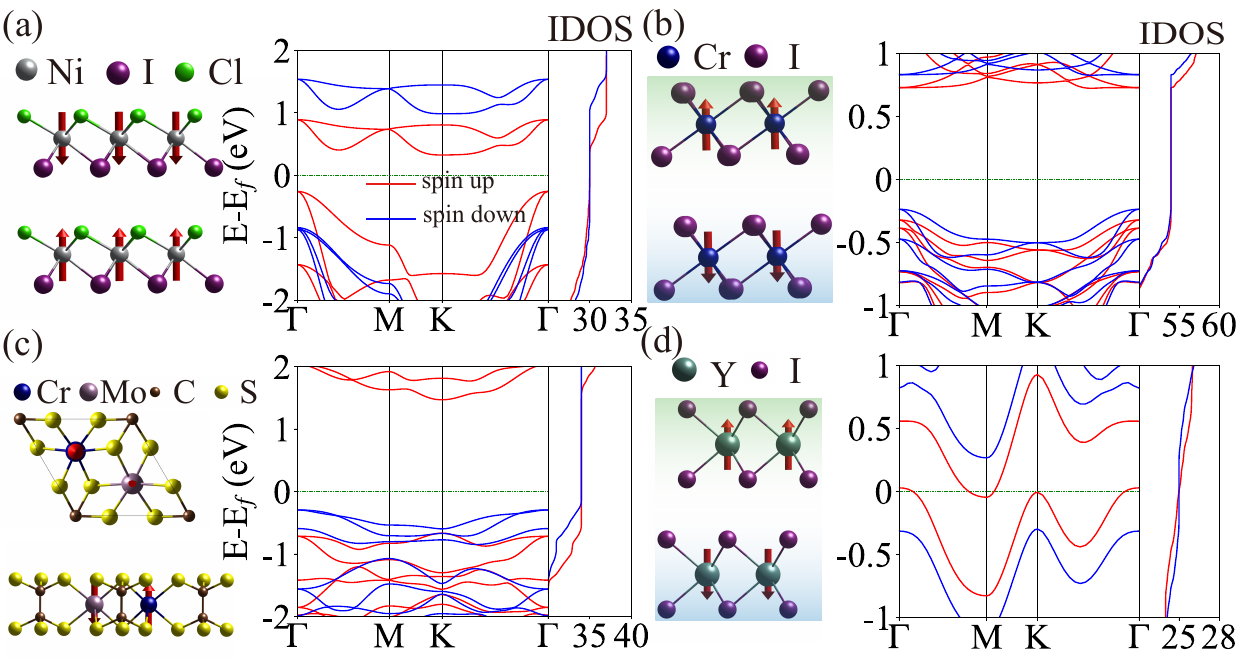}
    \caption{\label{fig:material}
Realization schemes and representative materials of 2D fully-compensated ferromagnetic materials. (a) The bilayer of NiICl with Janus structure and its band structure, as well as IDOS.  (b) The geometry and band structure of bilayer CrI$_3$ structure with a staggered potential of 0.5 eV. The staggered potential of the upper and lower layers is shown as the green and blue background. (c) The geometry and band structure of 2D CrMoC$_2$S$_6$, where the Mo atom replaces the Cr atom. (d) The geometry and band structure of YI$_2$ under 0.6 V/$\mathring{\text{A}}$ electric field. The system is half metal, and IDOS remains equal at Fermi energy with $N_\uparrow=N_\downarrow$, leading to a zero net magnetization. The geometry is plotted using VESTA software~\cite{momma2011vesta}}
\end{figure}

Interestingly, introducing the staggered potential ($\Delta\neq 0$) on sublattices breaks any connection between them by symmetry. Applying different onsite energies to the two sublattices can make them be considered composed of entirely different magnetic atoms or the same magnetic atoms under different environments. Consequently, no symmetry can connect the two sublattices, resulting in the spin splitting in bands, as shown in Figs.~\ref{fig:model}(d) and (e). Although the integrated densities of states (IDOSs) for two spin channels do not coincide across all energies, they are the same within the band gap of one spin channel with $N_{\uparrow}=N_{\downarrow}$, thereby keeping the total magnetization zero exactly. Notice that the \textit{filling-enforced} fFIM in Fig.~\ref{fig:model}(e) is a half metal, conducting only in one spin channel. Although the above discussion is based on the tetragonal lattice, our conclusions are general. We also build a model for the hexagonal lattice (see Supplementary Material (SM)~\cite{liusupplemental} for details).

\textit{Realizaiton schemes and materials.---}
To be an fFIM, a material must satisfy two conditions: (1) the number of occupied states with opposite spins must be equal, and (2) there should be no symmetry operations that connect the opposite spin sublattices. In this work, we propose three methods to construct fFIMs: (1) Janus structure, (2) staggered potential by external electric field or substrate, and (3) elemental substitution or alloying.

(1) Janus structure: The Janus structure breaks the symmetry connecting the two magnetic sublattices by altering the environment surrounding the magnetic atoms and induces a built-in vertical electric field. Such Janus structure is well known in the context of transition metal dichalcogenides~\cite{luJanusMonolayersTransition2017}. Here, we take the Janus structure NiICl as an example~\cite{guoLargeOutofplanePiezoelectric2022,gorkanSkyrmionFormationNibased2023}. It is derived by replacing the top layer I atoms in as-prepared NiI$_2$~\cite{songEvidenceSinglelayerVan2022} by Cl atoms with the inversion symmetry broken. The geometry and band structure of the bilayer NiICl are shown in Fig.~\ref{fig:material}(a). Although the Janus structure in the bilayer NiICl breaks the symmetry that typically restricts the system to zero net magnetization, as discussed above, as long as one of the spin channels is gapped, the total magnetic moment of the system remains zero, resulting in a fFIM.

(2) Staggered potential: By applying a staggered potential, introduced by an electric field or substrate, etc., to two magnetic sublattices, the symmetry connecting the two sublattices is broken, leading to spin splitting, and the system becomes a fFIM. From the above theoretical analysis, it can be seen that while a small staggered potential can induce gapped fFIMs, a large one can induce half-metallic fFIMs, as shown in Figs.~\ref{fig:model}(d) and (e). For example, in experimentally-prepared A-type antiferromagnetic bilayer CrI$_3$~\cite{huangLayerdependentFerromagnetismVan2017}, when a perpendicular electric field of 0.08 V/$\mathring{\text{A}}$ is applied, spin splitting occurs, and the spin-up and spin-down IDOSs remain equal in the gap, as shown in Fig.~\ref{fig:material}(b). In contrast, YI$_2$ under a perpendicular electric field of 0.6 V/$\mathring{\text{A}}$ is a half metal and exhibits gapless spin-up bands and gapped spin-down bands.  As described above, although the spin-up channel is gapless, as long as the spin-down channel is gapped, the system maintains $N_\uparrow=N_\downarrow$, and the spin-up integrated density of states (IDOS) is equal to the spin-down IDOS at the Fermi energy, as shown in Fig. \ref{fig:material}(d), demonstrating that the system remains fully compensated.

(3) Elemental substitution or alloying: By substituting or alloying with an element that either has the same number of valence electrons or differs by an even number compared with the original magnetic atom, ensuring these electrons are evenly distributed across both spin channels, the system maintains $N_\uparrow = N_\downarrow$, resulting in zero net magnetization. This method was first introduced in 3D systems~\cite{van1995half} and successfully implemented in Heusler alloys~\cite{stinshoffCompletelyCompensatedFerrimagnetism2017,semboshiNewTypeHalfmetallic2022}. As an example, we replace the magnetic Cr atoms in the 2D cAFM material CrCS$_3$ with Mo atoms from the same group in the periodic table, resulting in the material CrMoC$_2$S$_6$~\cite{wangTwoDimensionalQuaternaryTransition2022}. Despite maintaining zero net magnetization, the system exhibits ferromagnetic-like spin splitting, as illustrated in Fig.~\ref{fig:material}(c).

More details on the phonon spectra and magnetic ground states of bilayer NiICl, bilayer YI$_2$, and monolayer CrMoC$_2$S$_6$ are given in SM~\cite{liusupplemental}.
\begin{figure}[t]
    \centering
    \includegraphics[width=0.45\textwidth]{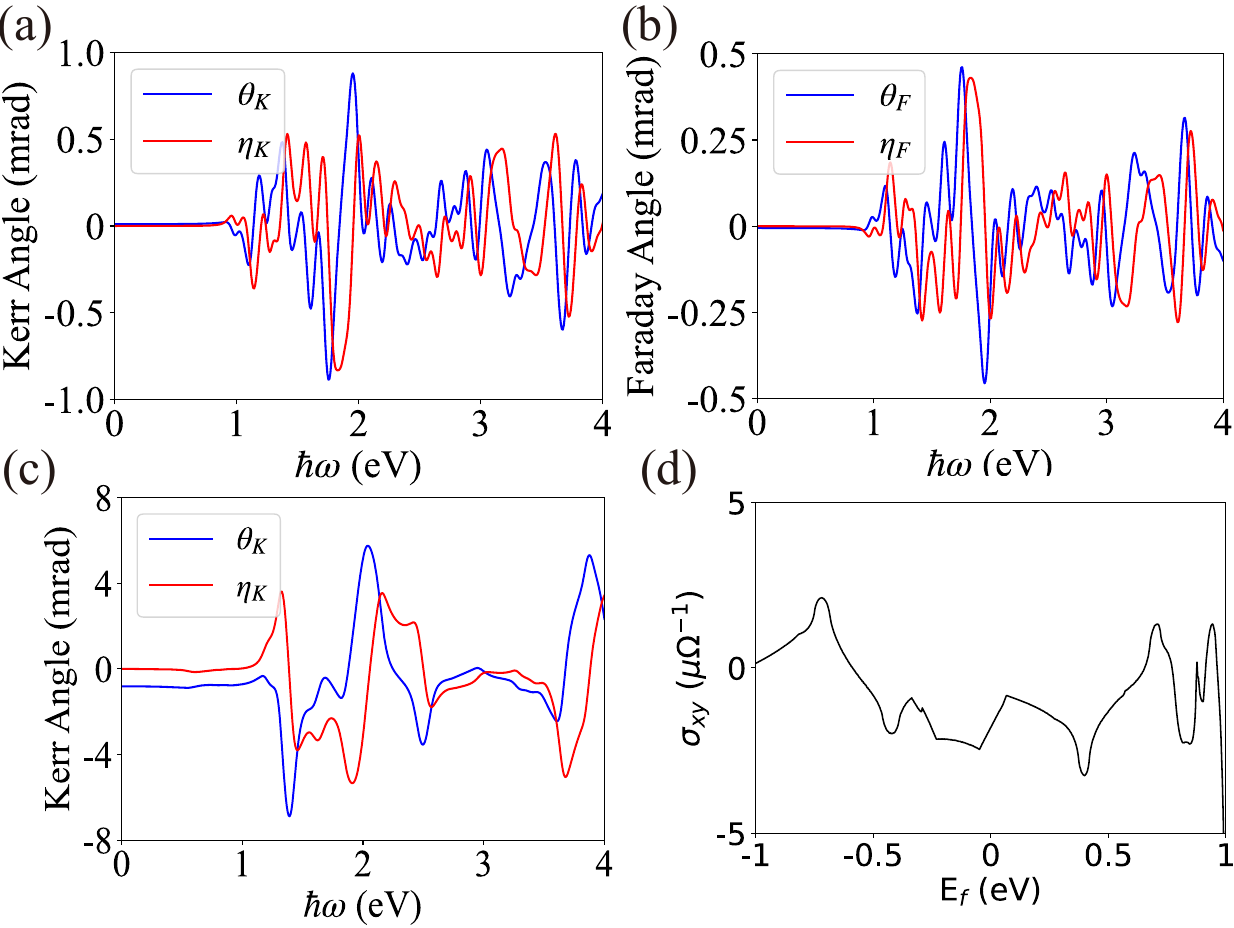}
    \caption{\label{fig:physical}Magneto-optical effect and anomalous Hall effect in 2D fully-compensated ferrimagnets. (a) and (b) show the Kerr angle and Faraday angle of fFIM bilayer CrI$_3$ with a 0.08 V/$\mathring{\text{A}}$ electric field. $\theta$ and $\eta$ represent the angle and ellipticity, respectively. (c) (d) The Kerr angle/AHC as a function of frequency/Fermi energy in half-metallic fFIM YI$_2$ with a 0.6 V/$\mathring{\text{A}}$ electric field.}
\end{figure}

\textit{Physical properties.---} fFIMs have zero net magnetization, making their magnetic properties similar to those of cAFMs and thus indistinguishable from cAFMs through external magnetic field measurements. However, our in-depth research reveals that fFIMs possess physical properties fundamentally different from cAFMs but more like ferromagnetic systems, such as the remarkable AHE, magneto-optical Kerr and Faraday effects, and fully spin-polarized currents, which are traditionally believed to occur only in ferromagnetic systems. To confirm these novel physical properties, we demonstrate our results in terms of both the effective model and representative materials. The above gapped fFIM CrI$_3$ and half-metallic fFIM YI$_2$ are taken as examples. The magneto-optical effect and AHE are also calculated by our fFIM tight-binding model in SM~\cite{liusupplemental}.

The magneto-optical Kerr and Faraday angles in 2D systems \cite{kimDeterminationInfraredComplex2007, valdesaguilarTerahertzResponseColossal2012}  are given by
\begin{equation}
\begin{split}
\theta_K+i\eta_K&=\frac{2(Z_0 d\sigma_{xy})}{1-(n_s+Z_0 d \sigma_{xx})^2},\\
\theta_F+i\eta_F&=\left(\frac{\sigma_{xy}}{\sigma_{xx}}\right)\left[1+\frac{n_s + 1}{Z_0 d\sigma_{xx}}\right]^{-1},\\
\end{split}
\end{equation}\label{Eq:MO}
where $\theta$ and $\eta$ represent the rotation angle and ellipticity, respectively. $\sigma_{xx}$ and $\sigma_{xy}$ are the diagonal and off-diagonal components of the optical conductivity tensor, quantifying the material's response to an applied electromagnetic field. $Z_0$ is the impedance of free space (approximately 377 $\Omega$), $d$ is the thickness of the simple, and $\epsilon_0$ is the vacuum permittivity. The refractive index of the substrate, denoted as $n_s$, is assumed to be 1.5, which corresponds to the refractive index of SiO$_2$.
The optical conductivity tensor $\sigma$ can be calculated by Kubo-Greenwood formula\cite{ebertMagnetoopticalEffectsTransition1996,aversaNonlinearOpticalSusceptibilities1995}
\begin{equation}
\sigma_{\alpha\beta}(\omega)=\frac{ie^2\hbar}{V}\int\frac{d^3 k}{(2\pi)^3}\sum_{n,m\neq n}\frac{f_{mn}}{\omega_{mn}}\frac{v_{nm}^\alpha v_{mn}^\beta}{\omega_{mn}-\omega}.\label{Eq:Kubo}
\end{equation}
Here $\omega_{mn}=\varepsilon_{m\bm k}-\varepsilon_{n\bm k}$ represents the energy difference between bands $m$ and $n$. $v_{nm}$ is the velocity operator, defined as $v_{nm}^\alpha=\bra{\psi_{n\bm k}}\partial_{\alpha} H_{\bm k}\ket{\psi_{m\bm k}}$. $\omega=\omega_0+i\eta$, where $\omega_0$ represents the frequency of the light and $\eta$ is the smearing parameter, set to 0.05 eV in this study.

While the pristine bilayer CrI$_3$ is cAFM, exhibiting no Kerr or Faraday effects, when a staggered potential is applied, the system is transited to an fFIM, with the resulting Kerr and Faraday angles shown in Figs.~\ref{fig:physical}(a) and (b), respectively. The maximum Kerr angle is close to 1 mrad, which is of the same order as that of monolayer CrI$_3$~\cite{huangLayerdependentFerromagnetismVan2017} and far exceeds the measurement limit of current equipment, demonstrating a significant magneto-optical effect.

With a large staggered potential, YI$_2$ exhibits half-metallic behavior, with only the spin-up channel being conductive, resulting in a significantly fully spin-polarized current. Furthermore, due to the lack of both space-time inversion symmetry and time-reversal symmetry, the system can display a nonzero AHC despite having zero net magnetization. The AHC is calculated using the formula~\cite{yaoFirstPrinciplesCalculation2004,wangInitioCalculationAnomalous2006}
\begin{equation}
\sigma_{xy} = \frac{-2 e^2}{\hbar V}\int\frac{d^2 k}{(2\pi)^3}\sum_{n,m\neq n}f_n\frac{\text{Im}[ v_{nm}^x(\bm k)v_{mn}^y(\bm k)]}{\omega_{mn}^2(\bm k)}.\label{Eq:AHC}
\end{equation}
The Kerr angle and AHC results for fFIM YI$_2$ are shown in Figs.~\ref{fig:physical}(c) and (d), and the Faraday angle is shown in SM~\cite{liusupplemental}. Despite zero net magnetization at the Fermi energy, fFIM YI$_2$ exhibits nonzero Kerr angle and AHC. The typical material results above and the general effective model calculations in SM demonstrate that 2D \textit{filling-enforced} fFIMs generally display a significant magneto-optical effect, and the half-metallic ones also show remarkable AHE and fully spin-polarized current.

\textit{Conclusion.---}  In summary, we propose a new type of 2D collinear magnets, distinct from cAFM and altermagnetism, termed 2D \textit{filling-enforced} fFIM, which achieves zero net magnetization through appropriate filling instead of symmetry. Thanks to being free of symmetry constraints, we put forward three universal schemes to realize 2D fFIMs: Janus structures, staggered potential, and elemental substitution or alloying. We demonstrate the effectiveness of these methods using NiICl with Janus structure, bilayer CrI$_3$ and YI$_2$ under an electric field, and CrMoC$_2$S$_6$ as examples, confirming their zero net magnetization and ferromagnetic-like band spin splitting through DFT calculations. We reveal that fFIMs display some physical properties that are similar to ferromagnets rather than cAFMs, such as magneto-optical effect, AHE, and fully spin-polarized currents. In contrast to altermagnets, the bands of \textit{filling-enforced} fFIMs have no spin degeneracy at the $\Gamma$ point. \textit{Filling-enforced} fFIMs are experimentally distinguishable from cAFMs, altermagnets, and ferromagnets through magneto-optical effect, spin-resolved ARPES, and direct magnetization measurements.

By utilizing the schemes presented in this work to achieve 2D \textit{filling-enforced} fFIMs, high-throughput screening can be employed to identify numerous candidate materials. These 2D fFIMs not only exhibit the physical properties of ferromagnetic materials but also possess the advantages of antiferromagnetic materials, along with the inherent tunability of 2D materials. Their remarkable physical properties, such as magneto-optical effect and AHE, as well as fully spin-polarized currents, will undoubtedly attract broad interest and are readily verified experimentally. The absence of stray magnetic fields in the 2D fFIMs allows for a significant increase in storage density, complemented by other benefits such as high response frequencies. All these merits indicate that our proposed 2D fFIMs will inject new and strong vitality into the development of spintronics.

\begin{acknowledgments}
\textit{Acknowledgments.}---The work is supported by the National Key R\&D Program of China (Grant No. 2020YFA0308800), the NSF of China (Grant No. 12374055), and the Science Fund for Creative Research Groups of NSFC (Grant No. 12321004).
\end{acknowledgments}

\bibliography{ref}

\end{document}


\title{Supplementary material for ``Two-dimensional fully-compensated Ferrimagnetism"}
\author{Yichen Liu$^{1}$}
\thanks{These authors contributed equally to this work.}
\author{San-Dong Guo$^{2}$}
\thanks{These authors contributed equally to this work.}
\author{Yongpan Li$^{1}$}
\author{Cheng-Cheng Liu$^{1}$}
\email{ccliu@bit.edu.cn}
\affiliation{$^{1}$Centre for Quantum Physics, Key Laboratory of Advanced Optoelectronic Quantum Architecture and Measurement (MOE), School of Physics, Beijing Institute of Technology, Beijing 100081, China}
\affiliation{$^{2}$School of Electronic Engineering, Xi'an University of Posts and Telecommunications, Xi'an 710121, China}
\maketitle
This supplementary material is organized into four sections. In the first section, we give a more detailed analysis of the square lattice model in the main text. In the second section, we demonstrate that 2D \textit{filling-enforced} fully-compensated ferrimagnets (fFIM) generally exhibit remarkable magneto-optical responses and anomalous Hall conductivity (AHC) by calculating these properties in a general tight-binding model. In the third section, we show how to convert traditional antiferromagnets into fFIM in a hexagonal model. The last section provides the detailed DFT calculations of NiICl, CrI$_3$, YI$_2$, and CrMoC$_2$S$_6$.

\section{SQUARE LATTICE TIGHT-BINDING MODEL}
In this section, we give a more detailed analysis of the square lattice model in the main text. The model in the main text is
\begin{align}
\begin{aligned}
H_\text{sq}=&t\sum_{i,d_j} (c_{1,i}^\dag c_{2,i+d_j}+ \text{h.c.})+\sum_{\alpha,i}(t_\alpha^x c_{\alpha,i}^\dag c_{\alpha,i+a_x}+t_\alpha^y c_{\alpha,i}^\dag c_{\alpha,i+a_y}+ \text{h.c.})+\Delta\sum_{\alpha,i} (-1)^\alpha c_{\alpha,i}^\dag c_{\alpha,i}+M\sum_{\alpha,i} (-1)^\alpha c_{\alpha,i}^\dag c_{\alpha,i} s_z.
\end{aligned}\label{Eq:Sq_model}
\end{align}
Here $t$ is the nearest hopping, and $t_1^x$, $t_2^x$, $t_1^y$, $t_2^y$ are the next-nearest hopping. $\Delta$ is the staggered potential, and $M$ is the magnetic moment. $d_j$ is the nearest vector, and $a_x$ and $a_y$ are the lattice vector. Then we transform the Hamiltonian into the reciprocal space and let $a_x=a_y=1$ to simplify the formula.
\begin{align}
H_\text{sq}=\left[(t_1^x+t_2^x)\cos k_x +(t_1^y+t_2^y)\cos k_y\right]\sigma_0 +4 t\cos\frac{ k_x}{2}\cos\frac{ k_y}{2}\sigma_x+\left[(t_1^x-t_2^x)\cos k_x+(t_1^y-t_2^y)\cos k_y+\Delta+Ms_z\right]\sigma_z.\label{Eq:Sq_AFM}
\end{align}
The bands have the following form
\begin{align}
\begin{aligned}
E_{\pm}^\uparrow=&(t_1^x+t_2^x)\cos k_x +(t_1^y+t_2^y)\cos k_y\pm\sqrt{\left[(t_1^x-t_2^x)\cos k_x+(t_1^y-t_2^y)\cos k_y+\Delta+ M\right]^2+4 t^2 (1+\cos k_x )(1+\cos k_y )},\\
E_{\pm}^\uparrow=&(t_1^x+t_2^x)\cos k_x +(t_1^y+t_2^y)\cos k_y\pm\sqrt{\left[(t_1^x-t_2^x)\cos k_x+(t_1^y-t_2^y)\cos k_y+\Delta- M\right]^2+4 t^2 (1+\cos k_x )(1+\cos k_y )}.\\
\end{aligned}\label{Eq:Sq_model_k}
\end{align}
The $\uparrow$ and $\downarrow$ in the superscript correspond to the spin-up and spin-down bands.

\begin{figure}[t]
    \centering
    \includegraphics[scale=0.5]{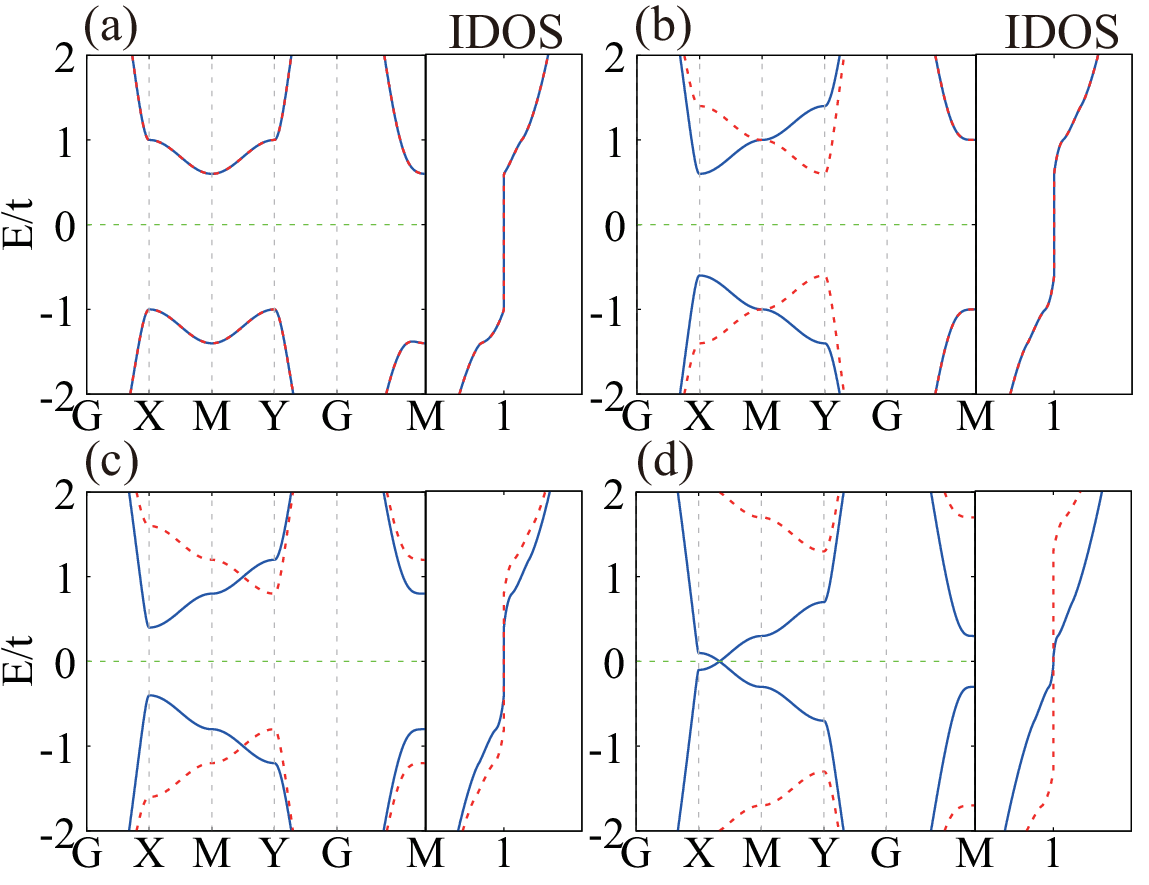}
    \caption{Tight binding model, with varying parameters yielding the conventional antiferromagnetism (cAFM), altermagnetism (AM), and fully-compensated ferrimagnetism models (fFIM). (a) The spin-degenerate band structure of the AFM model (Eq.~\ref{Eq:Sq_AFM}) with $t_1^x=t_2^x=t_1^y=t_2^y=0.1t$, $M=t$, and $\Delta=0$. The right panel shows the spin-degenerate spin-up (red) and spin-down (blue) IDOS. (b) The spin-splitting band structure of the AM model (Eq.~\ref{Eq:Sq_AM}), with parameters set as $t_1^x=t_2^y=0.1t$, $t_1^y=t_2^x=-0.1t$, $M=t$, and $\Delta=0$. The spin-up and spin-down IDOSs remain equal. (c) The spin-splitting band structure of the fFIM model ( Eq.~\ref{Eq:Sq_fFIM}) with parameters set as $t_1^x=t_2^y=0.1t$, $t_1^y=t_2^x=-0.1t$, $M=t$, and $\Delta=0.2t$. The spin-up and spin-down IDOS degeneracy is lifted but remains equal within the gap. (d) The half-metal band structure of the fFIM model, with parameters set as $t_1^x=t_2^y=0.1t$, $t_1^y=t_2^x=-0.1t$, $M=t$, and $\Delta=0.7t$. The spin-down channel is gapless, while the spin-up channel is gapped. The spin-up and spin-down IDOSs are equal at the Fermi energy.
    }\label{fig:sq_tb}
\end{figure}

When the parameters are set to $t_1^x = t_2^y = t_1^y = t_2^x = t_0$ and $\Delta=0$, the system maintains space-time inversion symmetry. Under this condition, the expression of energy bands are
\begin{align}
E_{\pm}^\uparrow=E_{\pm}^\downarrow = 2 t_0 (\cos k_x  + \cos k_y ) \pm\sqrt{M^2+4 t^2 (1 + \cos k_x )(1 + \cos k_y )},\label{Eq:Sq_AFM}
\end{align}
where energy bands are spin-degenerate, and the integrated density of states (IDOSs) are also spin-degenerate, as shown in Fig.~\ref{fig:sq_tb}(a).

When the parameters set to $t_1^x = t_2^y = t_1$, $t_2^x = t_1^y = t_2$ and $\Delta=0$, the parity-time symmetry is broken, but the $\{C_{2}||C_{4z}\}$ symmetry is remained, and the system becomes altermagnetism. The energy bands are given by
\begin{align}
\begin{aligned}
E_{\pm}^\uparrow = (t_1 + t_2) (\cos k_x  + \cos k_y ) \pm \sqrt{((t_1 - t_2)(\cos k_x  - \cos k_y ) + M)^2 + 4 t^2 (1 + \cos k_x )(1 + \cos k_y )},\\
E_{\pm}^\downarrow = (t_1 + t_2) (\cos k_x  + \cos k_y ) \pm \sqrt{((t_1 - t_2)(\cos k_x  - \cos k_y ) - M)^2 + 4 t^2 (1 + \cos k_x )(1 + \cos k_y )}.\\
\end{aligned}\label{Eq:Sq_AM}
\end{align}
It can be verified that in the energy band formula given by Eq.~(\ref{Eq:Sq_AM}), we can observe $E_{\pm}^\uparrow(k_y,-k_x)=E_{\pm}^\downarrow(k_x,k_y)$. This equation originates from the $\{C_{2}||C_{4z}\}$ symmetry, which transforms $k_x\to k_y$ and $k_y\to -k_x$ as well as spin up to spin down.
Therefore, the two bands with opposite spins are connected through the $\{C_{2}||C_{4z}\}$ symmetry. Thus the IDOSs maintain spin-degeneracy, as shown in Fig.~\ref{fig:sq_tb} (b).

Furthermore, when $t_1^x = t_2^y = t_1$, $t_2^x = t_1^y = t_2$ and $\Delta\neq 0$, the energy bands become
\begin{align}
\begin{aligned}
E_{\pm}^\uparrow = (t_1 + t_2) (\cos k_x  + \cos k_y ) \pm \sqrt{((t_1 - t_2)(\cos k_x  - \cos k_y ) +\Delta + M)^2 + 4 t^2 (1 + \cos k_x )(1 + \cos k_y )},\\
E_{\pm}^\downarrow = (t_1 + t_2) (\cos k_x  + \cos k_y ) \pm \sqrt{((t_1 - t_2)(\cos k_x  - \cos k_y ) +\Delta - M)^2 + 4 t^2 (1 + \cos k_x )(1 + \cos k_y )}.
\end{aligned}\label{Eq:Sq_fFIM}
\end{align}
The $C_{4z}$ symmetry is broken, there is no symmetry to connect sublattices of opposite spins, nor is there any symmetry to ensure zero total magnetization. However, if the number of spin-up occupied states ($N_\uparrow$) and spin-down occupied states ($N_\downarrow$) are equal, the total magnetization remains zero, resulting in a fFIM. The corresponding band structure and IDOS are shown in Fig.~\ref{fig:sq_tb} (c), where the spin-up and spin-down IDOSs are equal in the gap. If the energy bands for one spin channel remain gapped, the corresponding $N_\uparrow$ or $N_\downarrow$ stays a constant integer. Therefore, even if the other spin channel's bands close, as long as the total number of occupied states $N = N_\uparrow + N_\downarrow$ remains unchanged, $N_\uparrow = N_\downarrow$ still holds, maintaining the fFIM state. This situation, where one spin channel is metallic and the other is insulating, is known as a half-metal system. The band structure and IDOS are shown in Fig.~\ref{fig:sq_tb} (d), where the spin-up and spin-down IDOSs are equal at Fermi energy.

\begin{figure}[t]
    \centering
    \includegraphics[scale=0.5]{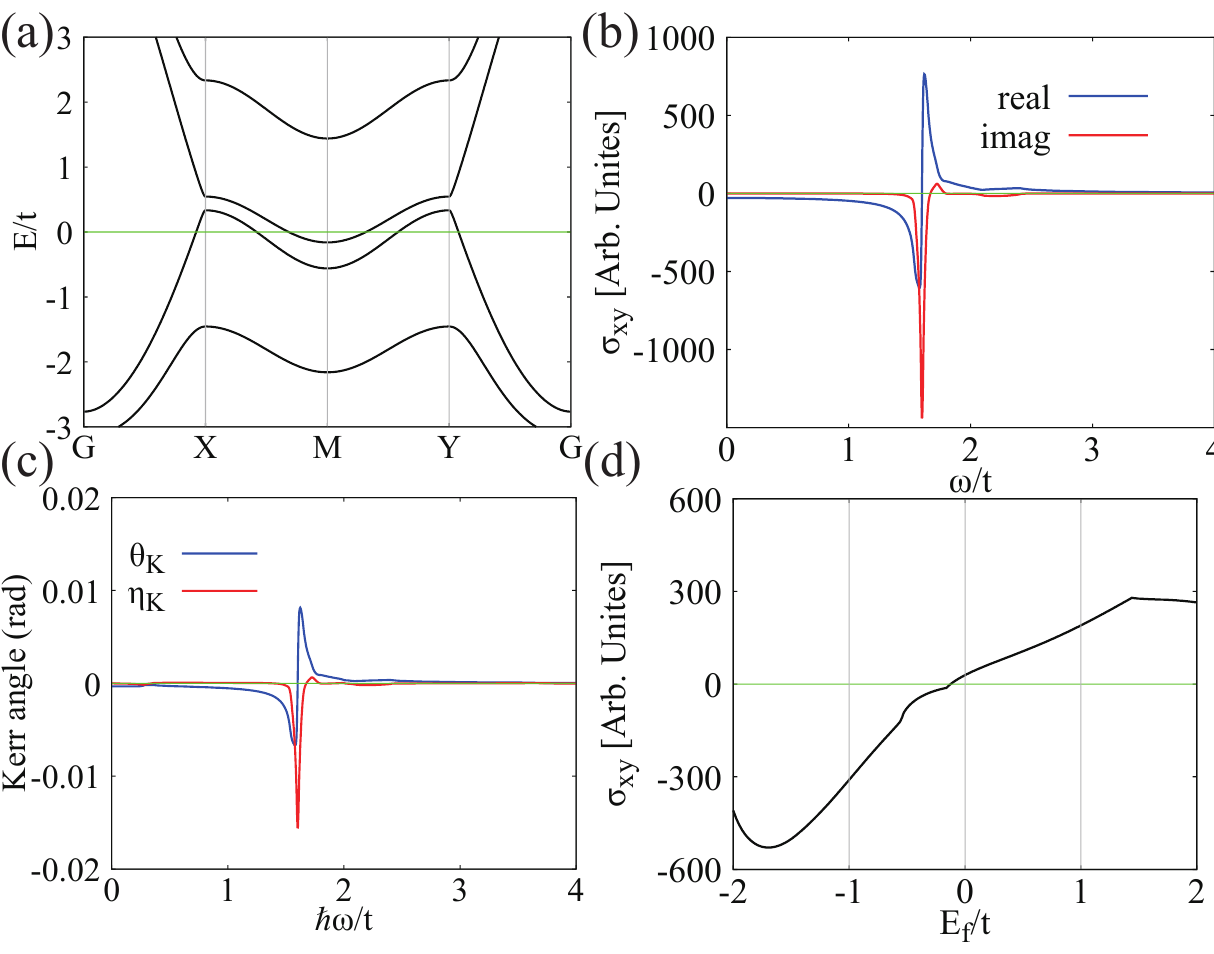}
    \caption{ The energy bands, Kerr angle, Faraday angle, and anomalous Hall conductivity of the general model of fully-compensated ferrimagnetism. (a) The band structures of this model, indicating metallic properties.  (b) The optical Hall conductivity as a function of the light frequency, where the red (blue) lines represent the real (imaginary) parts of $\sigma_{xy}$. (c) The Kerr angle as a function of the light frequency. $\theta$ and $\eta$ represent the angle and ellipticity, respectively. (d) The anomalous Hall conductivity as a function of the Fermi energy. The conductivity is nonzero at zero net magnetization. The parameters are fixed as $t_1^{x/y} = t_2^{x/y} = 0.2t$, $\Delta=0.7t$, and $\lambda = 0.2t$.
    }\label{fig:sq_soc}
\end{figure}

\section{The remarkable magneto-optical effect and anomalous Hall effect in general fFIM model}
Next, we further explore the novel physical properties of fFIMs based on this model. The Rashba SOC effect is introduced to the model, and the Hamiltonian can be written as
\begin{align}
H_\text{SOC}= H_\text{sq}+\sum_{i,d_j}\left[i\lambda(\bm d_j^0\times\bm s)_zc_{1,i}^\dag c_{2,i+d_j}+h.c.\right],\label{Eq:Sq_SOC}
\end{align}
with $\bm d_j^0=\bm d_j/|\bm d_j|=\{\frac{\sqrt{2}}{2}(1,1),\frac{\sqrt{2}}{2}(1,-1),\frac{\sqrt{2}}{2}(-1,1),\frac{\sqrt{2}}{2}(-1,-1)\}$. The Hamiltonian in reciprocal space is
\begin{align}
H_\text{SOC}=H_\text{sq}+2\sqrt{2}\lambda\left(\cos\frac{k_x}{2}\sin\frac{k_y}{2} \sigma_x s_x-\cos\frac{k_y}{2}\sin\frac{k_x}{2} \sigma_x s_y\right).
\end{align}
We focus on a metallic phase with $t_1 = t_2 = 0.2t$, $\Delta=0.8t$, $M=t$, and $\lambda = 0.2t$, as shown in Fig.~\ref{fig:sq_soc}. There are two hole pockets and one electron pocket in its band structure in Fig.~\ref{fig:sq_soc}(a). This model is free of space-time inversion symmetry, exhibiting ferromagnetic-like physical properties such as optical Hall effect, magneto-optical Kerr effect, and anomalous Hall effect. The calculation formula of optical Hall conductivity is~\cite{aversaNonlinearOpticalSusceptibilities1995,ebertMagnetoopticalEffectsTransition1996}
\begin{equation}
\sigma_{xy}(\omega)=\frac{ie^2\hbar}{V}\int\frac{d^3 k}{(2\pi)^3}\sum_{n,m\neq n}\frac{f_{mn}}{\omega_{mn}}\frac{v_{nm}^x v_{mn}^y}{\omega_{mn}-\omega}.\label{Eq:Kubo}
\end{equation}
Here $v^x$ and $v_y$ are the velocity operators, and $\omega_{mn}=\varepsilon_{m\bm k}-\varepsilon_{n\bm k}$ represents the energy difference between bands $m$ and $n$. $\omega=\omega_0+i\eta$, where $\omega_0$ represents the frequency of the light and $\eta$ is the smearing parameter, set to 0.05 eV in this study. The optical Hall conductivity $\sigma_{xy}$ of the model with Hamiltonian Eq.~(\ref{Eq:Sq_SOC}) is shown in Fig.~\ref{fig:sq_soc}(b). Furthermore, we can use the optical Hall two-dimensional system given by~\cite{kimDeterminationInfraredComplex2007,valdesaguilarTerahertzResponseColossal2012}
\begin{align}
\theta_K+i\eta_K&=\frac{2(Z_0 d\sigma_{xy})}{1-(n_s+Z_0 d \sigma_{xx})^2},\label{Eq:MO}
\end{align}
where $\theta$ and $\eta$ are the rotation angle and ellipticity, respectively. $\sigma_{xx}$ and $\sigma_{xy}$ are the diagonal and off-diagonal components of the optical conductivity tensor, quantifying the material's response to an applied electromagnetic field. $Z_0$ is the impedance of free space (approximately 377 $\Omega$), and $\epsilon_0$ is the vacuum permittivity. $d$ is the thickness of the sample and is set as 1 nm for the model. The refractive index of the substrate, denoted as $n_s$, is assumed to be 1.5, which corresponds to the refractive index of SiO$_2$. The Kerr angle of the model is illustrated in Fig.~\ref{fig:sq_soc}(c).

Since the system is metallic, we can also calculate AHC, whose formula is~\cite{yaoFirstPrinciplesCalculation2004,wangInitioCalculationAnomalous2006}
\begin{equation}
\sigma_{xy} = \frac{-2 e^2}{\hbar V}\int\frac{d^2 k}{(2\pi)^3}\sum_{n,m\neq n}f_n\frac{\text{Im}[ v_{nm}^x(\bm k)v_{mn}^y(\bm k)]}{\omega_{mn}^2(\bm k)}.\label{Eq:AHC}
\end{equation}
We calculate the AHC as a function of Fermi level $E_f$, as shown in Fig.~\ref{fig:sq_soc}(d). It is worth noting that when $E_f=0$, the model has zero net magnetization, but it has non-zero AHC.

\section{HEXAGONAL LATTICE TIGHT-BINDING MODEL}

\begin{figure}[t]
    \centering
    \includegraphics[scale=0.5]{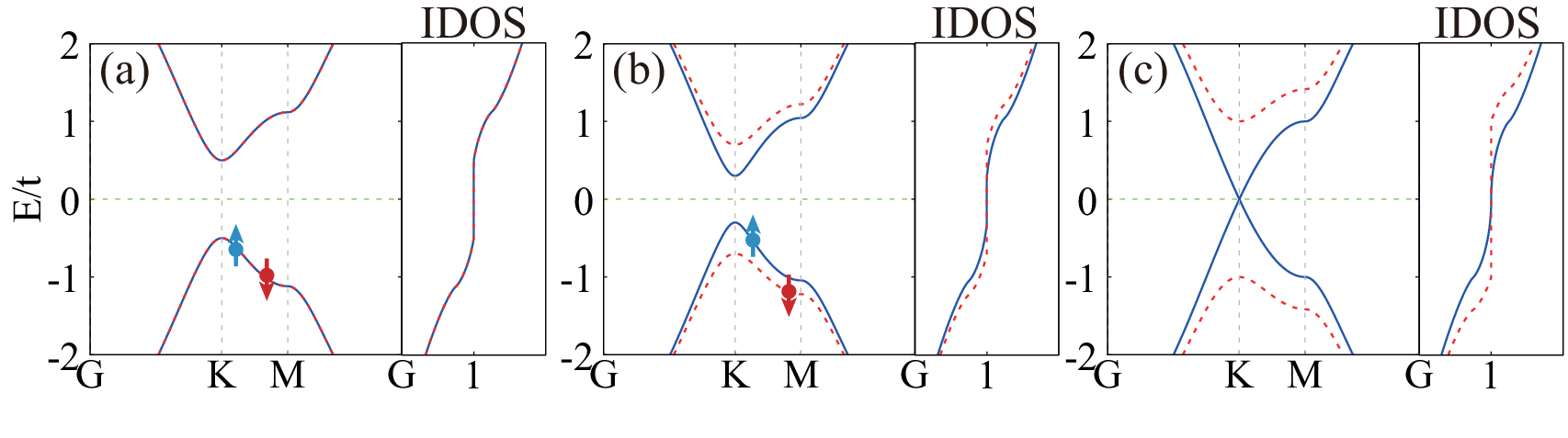}
    \caption{ Band structure and integrated density of states (IDOS) for the hexagonal model described by Eq.~(\ref{Eq:AFM}) with various parameter configurations. (a) The spin-degeneracy band structure of the model of Eq.~(\ref{Eq:AFM}). The right panel shows the spin-up (red) and spin-down (blue) IDOS. (b) The spin-splitting band structure of the model of Eq.~(\ref{Eq:Sq_fFIM}). The spin-up and spin-down IDOSs are degenerate in the gap. (c) The band structure of Eq.~(\ref{Eq:Sq_fFIM}) when $\Delta=M$. The system is a half metal.
    }\label{fig:tb}
\end{figure}

For the model in a hexagonal lattice, we begin by considering a hexagonal model with an antiferromagnetic term on the lattice
\begin{align}
H_\text{hex}=t\sum_{<ij>}c_i^\dag c_j+M s_z\sum_{i}(-1)^i c_i^\dag c_i.\label{Eq:AFM}
\end{align}
Here, $t$ represents the nearest-neighbor hopping, $M$ is the magnitude of the magnetic moment on each site, $s_z$ is the spin operator, and $(-1)^i$ represents an opposite sign of spins between sublattices. The two lattice is set as $\bm a_1=(1,0,0)$ and $\bm a_2=(-\frac{1}{2},\frac{\sqrt{3}}{2},0)$. In this model, the two sublattices possess opposite spins, and can be connected through inversion symmetry. Thus the system exists parity-time reversal symmetry. The energy bands of this model are given by
\begin{align}
E_\pm^{\uparrow\downarrow}=\pm\sqrt{3+M^2+2\cos k_x+4\cos\frac{k_x}{2}\cos\frac{\sqrt{3}}{2}k_y}.
\end{align}
Energy bands are spin degenerate, and spin-up and spin-down IDOSs are also degenerate, as illustrated in Fig.~\ref{fig:tb}(a). To break the inversion symmetry, we introduce a staggered potential
\begin{align}
H'=\Delta\sum_i (-1)^i c_i^\dag c_i.\label{Eq:fFIM}
\end{align}
Now, the two sublattices can no longer be related by any symmetry. Consequently, the energy bands of the model become
\begin{align}
\begin{aligned}
E_\pm^\uparrow=\pm\sqrt{3+(M+\Delta)^2+2\cos k_x+4\cos\frac{k_x}{2}\cos\frac{\sqrt{3}}{2}k_y},\\
E_\pm^\downarrow=\pm\sqrt{3+(M-\Delta)^2+2\cos k_x+4\cos\frac{k_x}{2}\cos\frac{\sqrt{3}}{2}k_y}.
\end{aligned}\label{Eq:hfFIM}
\end{align}
The band structure is shown in Fig.~\ref{fig:tb}(b) with spin splitting. As long as $\Delta < M$, the energy gap remains open, then the spin-up and spin-down IDOSs are equal in the gap. Therefore, the net magnetic moment maintains zero, and the system becomes a fFIM, as shown in Fig.~\ref{fig:tb}(b). When $\Delta=M$, the spin-down channel is gapless, and the system becomes a half-metal fFIM, as shown in Fig.~\ref{fig:tb}(c). Then as $\Delta>M$, the spin-down channel is gapped again, and the system returns to a fFIM.

\section{MATERIALS CALCULATION}\label{sec:material}

In this section, the calculation details for NiICl, CrI$_3$, YI$_2$, and CrMoC$_2$S$_6$ are provided.

The first-principles calculation of structure optimization and band structures are using Vienna \textit{ab initio} simulation package(VASP)\cite{kresseEfficientIterativeSchemes1996}. The generalized gradient approximation (GGA) with Perdew-Burke-Ernzerhof (PBE)\cite{perdewGeneralizedGradientApproximation1996a} realization is considered. We use the EFIELD, IDIPOL, DIPOL, and LDIPOL parameters in VASP to account for the electric field. The phonon calculations were performed using density functional perturbation theory as implemented in the PHONOPY code \cite{togoFirstprinciplesCalculationsFerroelastic2008}. The \textit{ab initio} tight-binding models are constructed using the WANNIER90 code~\cite{marzariMaximallyLocalizedGeneralized1997,mostofiWannier90ToolObtaining2008,souzaMaximallyLocalizedWannier2001} to calculate the magneto-optical response and anomalous Hall conductivity.

Bilayer CrI$_3$ has been synthesized~\cite{burchMagnetismTwodimensionalVan2018b,huangLayerdependentFerromagnetismVan2017} and is an interlayer antiferromagnetic material, the lattice parameter is $a=6.87\ \mathring{\text{A}}$, and the interlayer distance is 3.51 $\mathring{\text{A}}$. The Hubbard U is set as $U=2.65$ eV~\cite{cococcioniLinearResponseApproach2005} for the 3d orbital of Cr atom, and the Electric field is set as $0.08$ V/$\mathring{\text{A}}$ along negative $c$-axis. A $9\times9\times1$ k-mesh was utilized for these calculations.

NiICl belongs to a hexagonal lattice with the space group P3m1 and possesses $C_{3z}$ symmetry. After optimization, the lattice constant a is 3.49 $\mathring{\text{A}}$, and the interlayer distance is 3.38 $\mathring{\text{A}}$. A k-mesh of $11\times11\times1$ is employed for the calculations. To account for strong electron correlations in the Ni 3d orbitals, we applied the Hubbard U correction with $U=4$ eV in our DFT calculations.

Monolayer CrMoC$_2$S$_6$ is obtained by replacing the Cr atom with Mo atoms in the monolayer antiferromagnetic material CrCS$_3$. The Hubbard U of Mo and Cr elements are set as 4 eV. The $9\times 9\times 1$ k-mesh is used for lattice relaxation and self-consistent calculation.

\begin{figure}[t]
    \centering
    \includegraphics[scale=0.5]{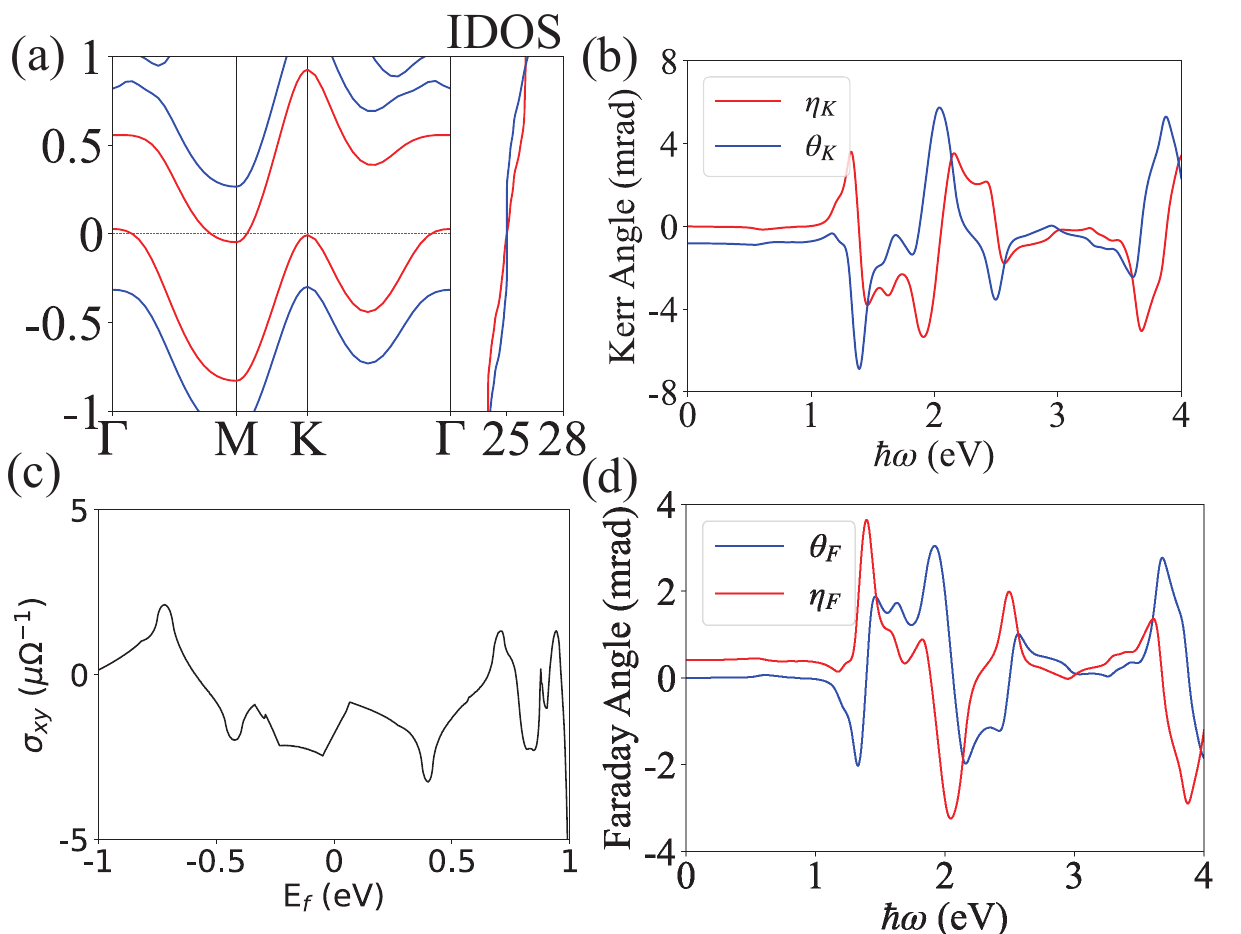}
    \caption{The band structure, Kerr angle, Faraday angle, and anomalous Hall conductivity of 2D \textit{filling-enforced} fully-compensated ferrimagnet YI$_2$. (a) Band structure and integrated density of states (IDOS). (b) The Kerr angle as a function of light frequency. $\theta$ and $\eta$ represent the angle and ellipticity, respectively. (c) The anomalous Hall conductivity as a function of Fermi energy $E_f$. (d) The Faraday angle as a function of light frequency.
    }\label{fig:YI2}
\end{figure}

AB-stacking bilayer YI$_2$ is an interlayer antiferromagnetic material, the lattice parameter is $a=4.10\ \mathring{\text{A}}$, and the interlayer distance is 3.42 $\mathring{\text{A}}$. The Hubbard U is set as $U=2$ eV for the 3d orbital of Y atom, and the electric field is set as $0.6$ V/$\mathring{\text{A}}$ along negative $c$-axis. A $9\times9\times1$ k-mesh is utilized for these calculations. The band structure, Kerr angle, Faraday angle, and AHC are shown in Fig.~\ref{fig:YI2}.

\begin{figure}[t]
    \centering
    \includegraphics[scale=0.9]{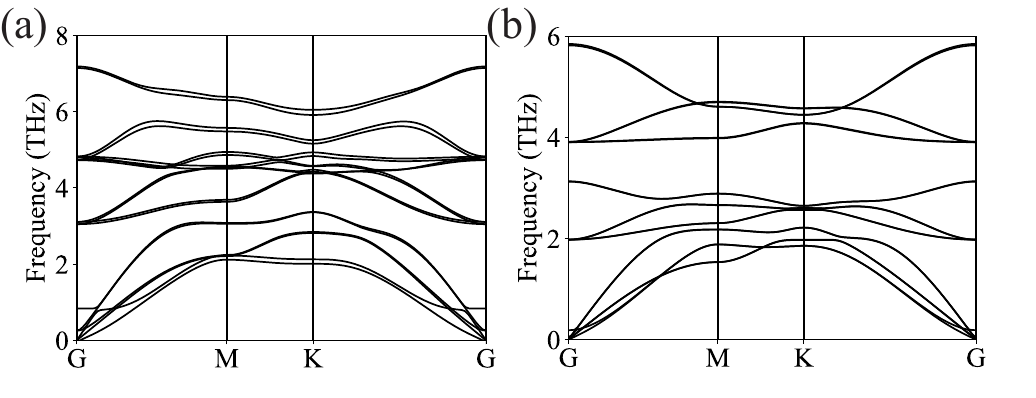}
    \caption{
The phonon spectra of bilayer NiICl (a) and bilayer YI$_2$ (b).}\label{fig:phonon}
\end{figure}

The stability of monolayer NiICl~\cite{gorkanSkyrmionFormationNibased2023}, monolayer CrMoC$_2$S$_6$~\cite{wangTwoDimensionalQuaternaryTransition2022} and monolayer YI$_2$~\cite{huangLargeSpontaneousValley2023} have been verified through DFT calculations on the phonon spectra.  Employing a $4\times 4\times 1$ supercell for bilayer NiICl and a $2\times2\times1$ supercell for bilayer YI$_2$, we calculate their phonon spectra, as shown in Fig. ~\ref{fig:phonon}. These spectra exhibit no imaginary frequencies in the Brillouin zone, indicating the structural stability of these materials.

The ferromagnetic and antiferromagnetic configurations of monolayer CrMoC$_2$S$_6$, bilayer NiICl, and bilayer YI$_2$ as a function of Hubbard U are listed in Table.~\ref{tab:table1} Within a reasonable range of the Hubbard U parameter, the ground states of CrMoC$_2$S$_6$, bilayer NiICl, and bilayer YI$_2$ are antiferromagnetic.

\begin{table}[t]
\caption{\label{tab:table1} The total energy of the ferromagnetic and antiferromagnetic order of CrMoC$_2$S$_6$ monolayer, NiICl bilayer, and YI$_2$ bilayer as a function of Hubbard U. E$_\text{AFM}$ and E$_\text{FM}$ are the total energy of the system with the antiferromagnetic and ferromagnetic order respectively. The E$_\Delta$ = E$_\text{AFM}$ - E$_\text{FM}$ is the total energy difference between the two magnetic orders. When E$_\Delta < 0$, the material is an antiferromagnet; otherwise, it is a ferromagnet. In the table, ``---'' indicates that VASP is unable to converge to the corresponding magnetic order. The Hubbard U of Cr and Mo are set as the same value in CrMoC$_2$S$_6$.}
\footnotesize
\begin{tabular}{|c|c c c c|}
\hline
\multicolumn{5}{|c|}{NiICl}\\ 
\hline
U (eV) & 2 & 3 & 4 & 5 \\
\hline
E$_\text{AFM}$ (eV) & -17.713227 & -16.865274 & -16.077713 & -15.348955 \\
E$_\text{FM}$ (eV) & -17.714349 & -16.864369 & -16.073680 & -15.341357 \\
E$_\Delta$ (eV) & 0.001122 & -0.000905 & -0.004033 & -0.007598 \\
\hline
\multicolumn{5}{|c|}{CaMoC$_2$S$_6$}\\ 
\hline
U (eV) & 1 & 2 & 3 & 4 \\
\hline
E$_\text{AFM}$ (eV) & -64.108138 & -62.529771 & -61.062467 & -59.686118 \\
E$_\text{FM}$ (eV) & --- & -61.917934 & -60.410827 & -58.999141 \\
E$_\Delta$ (eV) & --- & -0.611837 & -0.65164 & -0.686977 \\
\hline
\multicolumn{5}{|c|}{YI$_2$}\\ 
\hline
U (eV) & 0 & 1 & 2 & 3 \\
\hline
E$_\text{AFM}$ (eV) & -27.797280 & -26.962275 & -26.161623 & -25.395336 \\
E$_\text{FM}$ (eV) & -27.797266 & -26.962107 & -26.161330 & -25.394878 \\
E$_\Delta$ (eV) & -0.000014 & -0.000168 & -0.000293 & -0.000458 \\
\hline
\end{tabular}
\end{table}

\bibliography{ref}